\documentclass[letterpaper]{JHEP}

\usepackage{epsfig}
\usepackage{graphicx}

\preprint{SUGP-01/2-1\\
 hep-th/0103098}

\title{T-duality and the case of the disappearing brane}
\author{Donald Marolf\\Physics Department, Syracuse University, Syracuse,
         New York 13244}
\date{October, 2000}
\abstract{
When an NS5-brane crosses a D6-brane, a D4-brane is created stretching
between them by the Hanany-Witten effect.  However, the T-dual situation
involves a Kaluza-Klein monopole crossing a D5-brane and should not
result in brane creation.  Thus, the newly created D4-brane disappears
when T-duality is applied.  The T-duality is in a direction transverse
to the D4-brane so that one would naively have expected the creation of
a D5-brane in the T-dual picture.  This situation is investigated via the 
corresponding supergravity solutions and the tension with the naive
result is resolved.  Along the way, a differential form version of the
supergravity T-duality relations is developed and some comments are
made concerning the flux-expulsion properties of D6-branes.
}
\begin{document}
\section{Introduction}

In the last ten years it has come to be widely recognized that dualities 
are one of the fundamental structures of string- and M-theory.    
T-duality, which interchanges small circles with large ones, was
one of the first dualities to be discovered and understood, see e.g. \cite{Tdual}
for a review.  The familiar result that
applying T-duality in a direction
orthogonal to a Dp-brane results in a D(p+1)-brane (and similarly that 
T-duality applied along a Dp-brane results in a D(p-1)-brane)
is widely used to relate properties of the various Dp-branes.
It is therefore of interest to uncover any cases which deviate from
this rule and to carefully understand how such deviations occur.

The situation analyzed below is one to which precise application
of the above rule connecting
Dp- and D(p$\pm1$)-branes by T-duality would raise
a contradiction.  We consider
the case of a D4-brane created via the Hanany-Witten effect by 
the crossing of a D6-brane and an NS5-brane.  The two
branes have three spatial directions in common but are otherwise
orthogonal.  The D4-brane also lies along the three common directions but
its fourth direction is orthogonal to both the NS5 and D6-branes, since
the D4-brane stretches between them.  This setting is summarized by the
table below, where the X's indicate directions along the indicated brane.

\begin{center}
\begin{tabular}{|c|c|c|c|c|c|c|c|c|c|} \hline

{} & $x_1$ & $x_2$ & $x_3$ & $x_4$ & $x_5$ & $x_6$ & $x_7$ & $x_8$ & $x_9$ \\ \hline

D6 & X & X & X & X & X & X  &  &  & \\ \hline

NS5 & X & X & X  &  &  & & X & X  & \\ \hline

D4 & X & X & X &  & &  & & &  X \\ \hline

\end{tabular}
\end{center}

In the low
energy regime, the Hanany-Witten
effect \cite{HW}
can be understood from either the worldvolume (Born-Infeld-Chern-Simons)
theories of the branes \cite{wv} or from a spacetime 
perspective in supergravity \cite{3Q} as well as from a number of more
stringy arguments \cite{DFK,BGL,dA,HoWu,OSZ,NOYY,KZ}. 
We find the supergravity perspective
particularly useful in analyzing this situation and, as is discussed below, 
this analysis clearly shows the creation of the D4-brane. Such solutions were constructed
and analyzed in \cite{GM} based on the work of \cite{Aki,ITY}.  
Our setting is related to the more familiar creation of a fundamental
string by the crossing of D3 and D5-branes (see e.g. \cite{CGS})
by T-duality along the three
directions common to the NS5 and D6-branes and a final S-duality transformation
in the type IIB theory.

Consider, however, the application of 
T-duality in a direction along the D6-brane but
orthogonal to the NS5 and D4-branes (say along $x_4$).  This turns the NS5-brane into a
Kaluza-Klein monopole, so the IIB version is the crossing of a D5-brane
and a Kaluza-Klein monopole.  Naively, T-duality would turn the D4-brane
into a D5-brane.  However, one can show that D5-branes cannot
be created by the crossing of Kaluza-Klein monopoles and other D5-branes.
In fact, D5-branes can be created only in the presence of NS5-branes
or D7-branes.  

We find a resolution to this puzzle in the structure of the Kaluza-Klein
monopole.  The T-duality that creates the Kaluza-Klein monopole involves
a certain cutting and pasting of the original spacetime with the property
that closed surfaces in the original type IIA spacetime are not naturally
mapped to closed surfaces in the final IIB spacetime.  Ignoring this
fact would lead to the naive result, involving creation of D5-branes and
a contradiction.
However, a more careful treatment shows that no net D5 charge
is created and no Hanany-Witten effect takes place.
We analyze the situation using the corresponding supergravity solutions. 

The plan of the paper is as follows.  We begin with a brief review in section
\ref{shn} of
the picture
of the smooth creation of branes via the Hanany-Witten effect in supergravity.
It will be clear from this discussion that the
IIB intersection T-dual to the IIA 
D4-brane ending on the D6-brane cannot be merely a D5-brane three-brane junction
as one might otherwise expect.
In section \ref{IIB}, we use a differential 
form version of the supergravity
T-duality relations to map to the IIB setting, to
evaluate the various charges, and to illustrate the connection to
the closed surfaces mentioned above.  Finally, we end with some discussion
in section \ref{disc}, including comments relating D6-branes 
to ``superconducting black holes" \cite{sup}.

\section{The smooth picture of the Hanany-Witten effect}
\label{shn}

This section provides
a brief review of how the Hanany-Witten brane creation effect is described as
a smooth process.  The basic picture follows from general principles,  
but one can also find a one-parameter moduli space of exact 
BPS solutions describing certain
versions of the process in either the worldvolume theory of a 
D-brane in the background generated by another D-brane \cite{CGS} or in full
supergravity \cite{GM,3Q}.  What happens in
either the worldvolume theory or supergravity is that the flux of a gauge
field generated by one brane falling on the second brane generates a third kind
of charge associated with the new brane.  

In the supergravity description, this effect follows from the fact
that the `brane-source' charge of the D4-brane (see \cite{3Q}) is not
conserved \cite{GM,3Q}.  This in turn is a straightforward consequence
of the modified Bianchi identity satisfied by the gauge invariant
Ramond-Ramond  
four-form field strength $\tilde F_4$ of which the D4-brane is a magnetic
source.  We have the relation
\begin{equation}
\label{mbi}
d\tilde F_4 + F_2 \wedge H_3 = *j^{bs}_{D4}, 
\end{equation}
where the right hand side is the brane-source current (which vanishes
in the absence of an explicit D4-brane source).  Here, $F_2$ is the 
usual IIA Ramond-Ramond two-form field strength and $H_3 = dB_2$ is the
Neveu-Schwarz field strength.  Taking an exterior derivative of
(\ref{mbi}) shows that $d*j^{bs}_{D4}$ does not vanish.  Instead, 
a flux of $F_2$ falling on an NS5 brane (where $*j_{NS5}^{bs} \equiv
dH_3 \neq 0$) or a flux of
$H_3$ falling on a D6-brane (where $*j^{bs}_{D6} \equiv dF_2 \neq 0$) 
acts as a source or
sink of D4-brane charge. Some of the subtleties of defining charge and working
with brane-source currents are discussed in \cite{3Q}, but it is enough for
us that this result leads to the Hanany-Witten effect and the associated
creation of a D4-brane as described below.

The diagram below shows various stages in this process
for the case of an NS5-brane moving past a D6-brane to make a D4-brane
\cite{GM}. 
Similar results also follow for D$p$ and
D$p'$ branes whenever $p + p' = 8.$, see e.g. \cite{CGS} for a worldvolume
description of the D3/D5 case.    
At stage (i) when the NS5-brane is far from the D6-brane, 
the center of the
NS5-brane subtends a small angle at the D6-brane and captures only a small
amount of flux from the D6.  
As a result, essentially no D4 charge is induced in the region shown
and one has only
a flat NS5-brane.  Then, as the NS5-brane approaches the D6-brane (ii), 
it subtends a larger angle and begins to capture some flux, generating
some D4 charge.  This charge corresponds to D4-branes
lying inside the NS5-brane and running outward along this brane
to infinity.

\EPSFIGURE{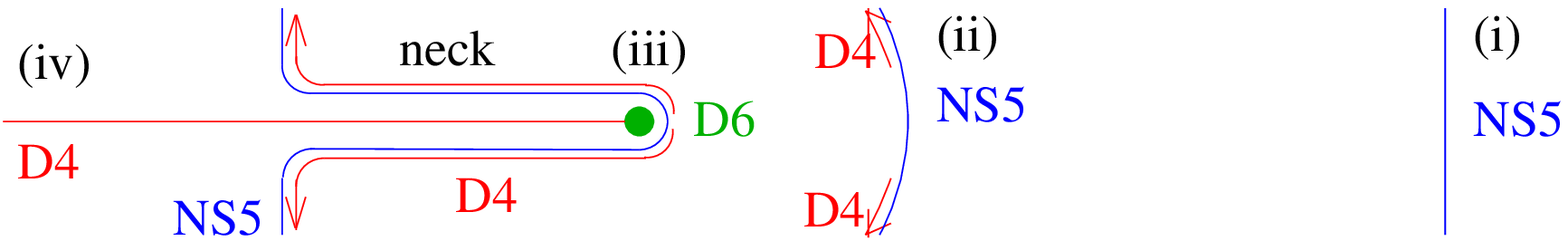,width=10cm}{Four stages in the Hanany-Witten process.}

When the NS5-brane is dragged past the D6-brane (iii), all
of the flux from the D6-brane is captured in the part of the NS5-brane close 
to the D6-brane.  Capturing one quantum of flux corresponds to the creation
of one quantum of fundamental string charge, so that the thin neck of
NS5-brane approximates a single D4-brane.  However, the NS5-brane
captures flux of the opposite sign in
the region where the neck joins the asymptotically flat part of
the NS5-brane.  The flux captured in this region is half of that
generated by the D6-brane, so that a net one-half quantum of D4 charge
reaches infinity along the NS5-brane.  This last statement is true
in each of the stages (i,ii,iii,iv), though only in stage (iii) are
all of the relevant parts of the NS5-brane visible in figure 1.  
In stage (iv), the neck has
narrowed so as to become difficult to resolve and all that remains is a D4-brane
string stretching between an NS5-brane and a D6-brane.  

One odd feature of the above discussion is that nowhere have we mentioned
the creation of D4-branes by $H_3$ flux falling on the D6-brane.
This is in fact the correct result, as may be seen from the explicit
exact solutions in \cite{GM}.  It is related to a `flux-expulsion'
effect associated with D6-branes and reminiscent of the `superconducting
branes' phenomenon \cite{sup}.  We will discuss this effect and related issues further
in section \ref{disc}.

Each stage in figure 1 depicts a BPS configuration
of either the worldvolume theory (as described in \cite{CGS}) or supergravity
(following \cite{GM}).
Studying BPS configurations is useful for a variety of well-known reasons.
However, one should also apply the lessons
learned from BPS configurations to non-BPS contexts.  For example, 
we learn from the above diagram that there is no fundamental difference
between a D4-brane ending on a D6-brane and an NS5-brane tightly
wrapped around the D6-brane.  Indeed, we would generically expect
fluctuations (dynamical, thermal, or quantum mechanical) between such
configurations.  For the case above, both configurations are BPS.
However, this is a function of the detailed boundary conditions: note
that the asymptotics\footnote{In both cases, the asymptotics
should be those of a logarithmic curve \cite{Aki1} 
due to the fact that the end of the D4-brane
in the NS5 is of co-dimension 2.  The case of a D2-brane
crossing a D6-brane and creating a fundamental string is very much
the same.  See \cite{half} for some interesting
comments on the relation of this logarithmic curve to the fundamental
string charge flowing through the D2-brane.  These comments also
apply to the current NS5/D6/D4 case.}
of the NS5-brane are different at stages (i) and (iv).  

If one fixes
the boundary conditions then there is only one BPS configuration and any
fluctuations should cost energy.  In particular, suppose that we fix the
boundary conditions to be those of stage (iv).  We then conclude that
there are fluctuations in which the attachment of the D4 to the D6
`loosens' into the configuration shown below\footnote{Again, the
same effect arises in a D2/D6/F1 system, with the fluctuations now
being between a tightly attached fundamental string and one loosely
attached by a loop of D2-brane.  One has to
wonder about the implications of this observation for understanding black
hole entropy non-perturbatively.  Naive extrapolation suggests that
at finite $g$ (and finite $N$) the states counted by Strominger and Vafa
\cite{SV} are not
well localized on D-branes.  Perhaps this is related to the fact
that black hole solutions of supergravity have interiors of finite size?}.
However, due to the tension
of the NS5-brane, this loosely attached configuration has more energy than 
the tightly attached configuration and the `noose' will tighten of its
own accord.  In particular, in the limit in which
string perturbation theory becomes valid ($g \rightarrow 0$ with $\ell_s$
fixed) the NS5-brane (whose tension is proportional to $1/g^2$)
becomes very heavy relative to a D4-brane (with tension $T_{D4} \sim 1/g$)
and the energy difference between the loosely and tightly attached configurations
diverges.  This is of course consistent with the 
usual observation that only the tightly attached configuration (of either
D4's or F1's ending on D6's) is relevant
in perturbation theory and that one sees no sign there of fluctuations
to a loosely attached configuration.

\EPSFIGURE{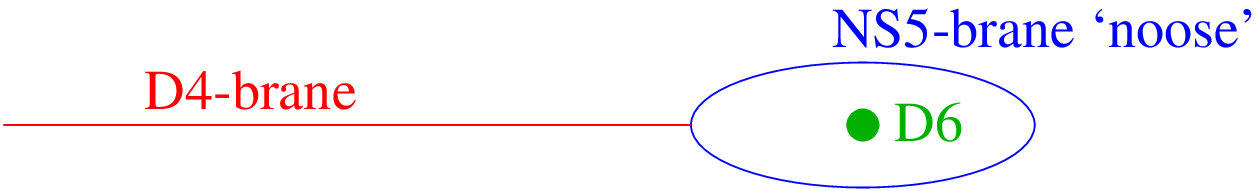,width=10cm}{A D4-brane `loosely attached'
to D6-brane.}

The above discussion followed directly from an analysis of D4 brane-source
charge conservation.  Let us now consider a T-dual configuration, where
the T-duality is applied in a direction along the D6-brane but
orthogonal to the NS5-brane.  This direction will also be orthogonal
to the D4-brane created above.

It is simplest to begin with stage (i), which contains only two widely
separated branes.  The T-dual configuration clearly contains a D5-brane
and a Kaluza-Klein monopole, which is the T-dual of the NS5-brane. 
To see whether a D5-brane can be created in stages (ii) and (iii), we
should analyze conservation of the D5 brane-source charge.  The
associated gauge invariant field strength $\tilde F_3$ does satisfy
a modified Bianchi identity, but in the form
\begin{equation}
\label{D5mbi}
d \tilde F_3 - F_1 \wedge H_3 = * j^{bs}_{D5},
\end{equation} 
where $F_1$ is the IIB one-form field strength associated with D7-branes.
Here our conventions for the IIB fields differ from those of \cite{joe} by a sign, in order
that we make a choice compatible with the conventions of \cite{Bus,MO,dielec}.
Taking an exterior derivative, we see that $*j^{bs}_{D5}$ is conserved
except perhaps at D7-branes or NS5-branes -- neither of which appear
in our setting.  As a result, D5 brane-source charge is conserved
and no creation of D5-branes can occur\footnote{To add further weight to
the argument it is
useful to notice that one could further S-dualize our IIB setting and
obtain a pure NS solution involving an NS5-brane and a Kaluza-Klein 
monopole.  One is then free to reinterpret this as a IIA solution and
lift to eleven dimensions.  In that context, the created brane
would be an M5-brane.  But the M-theory 4-form field strength satisfies
an unmodified Bianchi identity and M5 brane-source charge is 
conserved in all settings.}.  

Considering the ending of the D4-brane on the D6-brane, one might
have thought that the T-dual description should be a three D5-brane 
junction, which does indeed conserve
D5-brane charge: the charge simply flows from one brane
to the other.  However, this option is ruled out by the fact that on the
IIA side the D4-brane is smoothly created before the NS5 and D6-branes come
into contact.  As a result, there is no way for D5-brane charge
to flow between the branes in stages (ii) and (iii) above.

\section{The IIB description}
\label{IIB}

In this section, we analyze the charges of the T-dual IIB spacetime.
Luckily, we will find that one need not explicitly
write down the IIB solution.  Instead, we will be able to evaluate the 
various charges directly from our knowledge of the IIA case.

The charge in which we are particularly interested is the D5-charge
of the created brane.  Since the D5-charge is conserved in this setting, 
such charges can best be computed by integrating the
flux of $\tilde F_3$ over an appropriate sphere.  It is therefore
useful to work with the T-duality relations between IIA and IIB supergravity
written in a geometric way associated with differential forms 
instead of the original component descriptions
used in \cite{Bus,MO}.  We therefore briefly reformulate
these relations in subsection \ref{moreG} before moving on to study
the IIB solution in subsection \ref{study}.

\subsection{A more geometric form of the T-duality relations}
\label{moreG}

One can check that the results given in \cite{Bus} for the action of T-duality on
Neveu-Schwarz fields and in \cite{MO} for the action on Ramond-Ramond fields
are equivalent to the following description of
T-duality in IIA/IIB supergravity.  
Let us begin with a spacetime having a 
Killing field $\lambda$ with $S^1$ orbits.  Let
$\tilde \lambda_\mu = \lambda_\mu/|\lambda|^2$
denote the one-form obtained by lowering the index
on the Killing field $\lambda$ with the string metric and
dividing by its norm.  Let
$y$ denote any coordinate along the orbits of $\lambda$ satisfying
$\lambda^\mu \partial_\mu y =1.$  We normalize the Killing field by requiring
$y$ to have period $2\pi \ell_s$, where $\ell_s$ is the string length.  Thus, 
for a spacetime that may be treated classically $|\lambda|$ will be of size
$R/\ell_s$ for some macroscopic length scale $R$.

Just as in Kaluza-Klein reduction,
one can split the metric into a part ($ds^2_r$, the reduced metric)
associated with intervals transverse to the orbits of $\lambda$ and a
part along $\tilde \lambda:$
\begin{equation}
ds^2_{string} = ds^2_{r} + |\lambda|^2 \tilde \lambda \otimes \tilde
\lambda.
\end{equation}  
Similarly, the Neveu-Schwarz B-field can be reconstructed from the one-form
$(B \cdot \lambda)_\nu = B_{\nu \mu} \lambda^\mu$ and the two-form
$B^r = B - \frac{1}{2} (dy + B \cdot \lambda) \wedge \tilde \lambda.$
Introducing the general notation $(C_n \cdot \lambda)_{\alpha_1...
\alpha_{n-1}} = (C_n)_{\alpha_1...\alpha_{n-1}\alpha_n} \lambda^{\alpha_n}$
for an n-form $C_n$,
the reader may check that the T-duality relations of \cite{Bus,MO}
may be written:

\begin{eqnarray}
\label{geom}
\widehat{\phi} &=& \phi - \ln(|\lambda|) \cr
\widehat{ds}_r^2 &=& ds_r^2 \cr
\widehat{B}_r &=& B_r \cr
\widehat{\tilde \lambda} &=& dy + B \cdot \lambda \cr
dy + \widehat{B} \cdot \lambda &=& \tilde \lambda \cr
|\widehat{\lambda}|^2 &=& \frac{1}{|{\lambda}|^2} \cr
\widehat{C}_n &=& C_{n+1} \cdot \lambda
+ ( C_{n-1} - C_{n-1} \cdot \lambda \wedge \tilde \lambda)
\wedge (dy + B \cdot \lambda),
\end{eqnarray}        
where hatted symbols (e.g. $\hat{\tilde \lambda}$) denote fields in the T-dual
solution and $C_n$ are the Ramond-Ramond gauge potentials.  
The understanding in (\ref{geom}) is that the two solutions
are thought of as living on the same manifold\footnote{This manifold should consist
only of those points where both the metric and the NS gauge potential $B$ are
non-singular.} with $\lambda$ representing
the same vector field in each case.  However,
after performing the T-duality it may be natural to extend the manifold
to a larger one that is geodesically complete, e.g. in the
duality between the NS5-brane and the Kaluza-Klein monopole in which
the fixed points of the Killing field must be added by hand after applying
(\ref{geom}).  Note that although $\lambda^\mu$ represents the same vector field
on both the IIA and IIB sides, the one-forms $\tilde \lambda_\nu = 
\frac{\lambda^\mu g_{\mu \nu}}{\lambda^\alpha g_{\alpha \beta} \lambda^\beta}$
and $\widehat{\tilde \lambda}_\nu = 
\frac{\lambda^\mu \hat g_{\mu \nu}}{\lambda^\alpha \hat 
g_{\alpha \beta} \lambda^\beta}$ differ due to changes in the metric.

Since T-duality is properly applied when the orbits of the Killing vector
field are compact, one should think of these relations as holding on
a circle bundle over the reduced space of orbits.
Note that (\ref{geom}) is not completely geometric due to the
fact that T-duality mixes the metric and Neveu-Schwarz B-field and thus
mixes diffeomorphisms with NS gauge transformations.  It is for this
reason that a choice of coordinate $y$ must be made before one can write down
the above relations.  Of course, two different choices of $y$ result in
spacetimes that differ only by a Neveu-Schwarz gauge transformation. 

What we will need most to compute charges is not the
potentials given in (\ref{geom}) above, but the associated field strengths.
If we can choose a gauge in which the Lie derivatives of the potentials
$C_n, B$ vanish along the Killing field $\lambda$, then
the relations $d(C_n \cdot \lambda) = (dC_n ) \cdot \lambda$ hold
and the field strengths $F_n =dC_{n-1}$ satisfy

\begin{eqnarray}
\label{fs}
\widehat{F}_n &=& F_{n+1} \cdot \lambda
+ ( F_{n-1} - F_{n-1} \cdot \lambda \wedge \tilde \lambda + 
(-1)^{(n-2)} C_{n-2} \cdot \lambda \wedge d\tilde \lambda)
\wedge (dy + B \cdot \lambda) \cr
&+& (-1)^{n-2} ( C_{n-2} - C_{n-2} \cdot \lambda \wedge \tilde \lambda)
\wedge H \cdot \lambda,
\end{eqnarray}
where $H =dB$. 
Of course, not all of these field strengths are gauge invariant.
In particular, neither the field strength $F_4 = dC_3$ 
which couples magnetically to D4-branes nor the $F_3 = dC_2$ 
which couples magnetically to D5-branes is gauge invariant.
Instead, the gauge invariant versions are\footnote{Again, our conventions
are compatible with \cite{Bus,MO,dielec}.  However, while they agree with
those of \cite{joe} on the IIA side, they differ with \cite{joe} by a sign on the
IIB side.}
\begin{eqnarray}
\label{gifs}
{\tilde F}_3 &=& F_3 + C_0 \wedge H, \cr
{\tilde F}_4 &=& F_4 - C_1 \wedge H.
\end{eqnarray}
It follows from (\ref{fs}) above that these are related by
\begin{equation}
\label{f3}
\widehat{\tilde F}_3 = 
{\tilde F}_4 \cdot \lambda
+ ( F_2 - F_2 \cdot \lambda \wedge \tilde \lambda)  
\wedge (dy + B \cdot \lambda). 
\end{equation}

Taking a further exterior derivative leads to the conclusion that
the D4, D5, and D6 brane-source charges satisfy
\begin{equation}
*j_{D5}^{bs} = *j_{D4}^{bs} \cdot \lambda + (*j_{D6}^{bs} - *j_{D6}^{bs}
\cdot \lambda \wedge \tilde \lambda) \wedge (dy + B \cdot \lambda).
\end{equation}
Similar relations hold for the other D-branes.

As a result, at least if the branes appear in a part of the spacetime where
the circle bundles are smooth, the T-dual
of a D4 brane-source current is indeed a D5 brane-source current when the T-duality
acts orthogonally to the D4-brane.
 
\subsection{Surfaces, Charges, and disappearing branes}
\label{study}

From the results of the last subsection, one would be tempted to conclude
that our D4-branes are replaced by D5-branes in the
T-dual type IIB spacetime.  However, the subtlety is that these branes 
would lie exactly at the points where the circle bundle description breaks
down.  In particular, on the
the IIB side they would lie at the core of the Kaluza-Klein
monopole where the circles degenerate to a point\footnote{Singularities
associated with branes can sometimes be removed by smearing out the charge
over a finite volume.  However, there is no smooth spacetime corresponding
to a smeared Kaluza-Klein monopole just as there is no fiber bundle that 
realizes the idea of a smeared magnetic charge.}.  As a result, 
we must compute D5-brane charge by considering flux integrals of ${\tilde F}_3$
over various surfaces in the spacetime.  Since $\lambda$
is along the D6-brane, one expects $C_1\cdot \lambda$ and
$F_2 \cdot \lambda$ to vanish on the IIA side.  This may be
checked explicitly using the solution of \cite{GM}.
One may also check that the Lie derivatives along $\lambda$ of $C_n$ and $B$ vanish.
As a result, equation (\ref{f3}) becomes
just

\begin{eqnarray}
\label{f32}
\widehat{\tilde F}_3 &=& 
{F}_4 \cdot \lambda - C_1 \wedge H \cdot \lambda
+  F_2 
\wedge (dy + B \cdot \lambda) \cr
&=& d \left[
{C}_3 \cdot \lambda +
 C_1 
\wedge (dy + B \cdot \lambda) \right]. 
\end{eqnarray}
In particular, $\widehat{\tilde F}_3$
is a total derivative, as one would expect in a situation
in which the associated D5-brane charge is conserved.  As a result, 
integrals of this field over a 
closed surface can be evaluated purely in terms of contributions
from submanifolds where the potentials are ill-defined; i.e., from Dirac strings.   
It is useful to note that our D5 charge can be evaluated
from the Dirac strings of the associated type IIA fields $C_3$,
$C_1$, and $B$. 

\EPSFIGURE{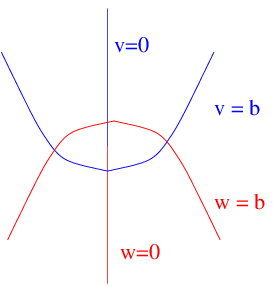,width=4cm}{The v,w coordinates.}

Before proceeding, we pause briefly to acquaint the reader with the coordinate
system used in \cite{GM}.  The Cartesian coordinates $x_0$ (time), and
$x_1,x_2,x_3$ are taken to run along all three branes (D6, NS5, and D4).
The T-duality will be performed along the $x_4$ direction, which  
is also along the D6-brane.  Here consider the
D4- and D5-branes to be smeared along this direction in order to generate
the required translational symmetry.  The coordinates $x_5,x_6$ are
along the D6-brane but transverse to the NS5- and D4-branes.

For the rest of the coordinate system, we take $x_7,x_8,x_9$ to be
Cartesian-like coordinates in the space transverse to the D6-brane which
are defined in a manner adapted to the spherical symmetry of the $F_2$ field. 
When the NS5-brane is far from the D6-brane (i.e., stage (i)), we may think
of it as lying in the $x_7,x_8$ directions at $x_9 = constant$.  However, 
when the NS5-brane approaches the D6-brane it deforms and is no longer
flat.  As a result,
to describe stages (ii) through (iv)
it is useful to introduce the coordinates $v, w$ through
\begin{eqnarray}
r^2 &=& x_7^2 + x_8^2 + x_9^2, \cr
\cos \theta &=& x_9/r, \cr
v &=& r \sin \theta/2, \cr
w &=& r \cos \theta/2,  
\end{eqnarray}
In particular, in the region close to the D6-brane core the NS5-brane
lies on the surface $v= constant$.

\EPSFIGURE{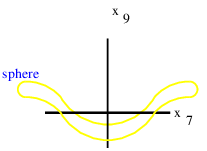,width=5cm}{The upper sphere.}

We are interested in the following surfaces.  First, we wish
to measure the total D5-brane charge flowing down from infinity
along the Kaluza-Klein monopole.  So, we need a 3-sphere surrounding
the $v=0$ axis (and in fact outside the surface $v=b, x_5=x_6=0$).  
One particular choice in the IIA near D6-brane spacetime
would be the surface $S_U$ ($U$ for upper) defined by
\begin{eqnarray}
\label{usphere}
v^2 + w^2 \equiv r^2 = const, \cr
v^2 + x_5^2 + x_6^2 \equiv R_U^2 = const, \cr
x_0,x_1,x_2,x_3,x_4=0.
\end{eqnarray}
This three-sphere lies above the second brane and surrounds the
monopole so long as $r^2 > 2R_U^2$ and $R_U > b$.  A `side view'
of this sphere is shown below, where the $x_7$ and $x_9$ axes are shown
and the direction out of the page represents, say, $x_6.$  The closer side of
the sphere passes in front of the $x_7$ and $x_9$ axes with the farther
size passing behind.

We would also like to measure the charge flowing outward along the second
brane.  To do so requires a surface surrounding this brane in an
appropriate sense.  Some thought shows that the correct surface has
the topology $S^2 \times S^1$ and that a particular example is the surface
$S_0$ given
by taking 
\begin{eqnarray}
\label{0sphere}
v^2 + w^2 \equiv r^2 = const, \cr
x_5^2 + x_6^2 \equiv R_0^2 = const, \cr
x_0,x_1,x_2,x_3,x_4=0.
\end{eqnarray}

To double check the results we fill find, we might also consider a surface lying entirely
below the second brane.  The flux through this sphere should be
the sum (with an appropriate sign) of the fluxes through the two surfaces
described above.
The sphere $S_L$ defined by 
\begin{eqnarray}
\label{lsphere}
v^2 + w^2 \equiv r^2 = const, \cr
w^2 + x_5^2 + x_6^2 \equiv R_L^2 = const, \cr
x_0,x_1,x_2,x_3,x_4=0,
\end{eqnarray}
lies below the second brane for $r^2 > 2 R_L^2$. 

Let us now consider the Dirac strings of the IIA fields.
As in \cite{GM}, we may
we work in a gauge where the Dirac string of $C_1$ lies at $v=0$ and
does not intersect any of the surfaces above.  From \cite{GM} one finds
that such a gauge has
$C_3=0,$ so that Dirac strings of $C_3$ are not a concern.

What remain are only the IIA Dirac strings of $B$.  We pause here to
note that these become Dirac strings of the Kaluza-Klein monopole in
the IIB solution.  At a Dirac string of $B$, one must perform a large
gauge transformation in order to smooth out the description of the field.
For the Kaluza-Klein monopole, the associated smoothing requires
a large coordinate transformation.
The result is that if a Dirac string of $B$ intersects a smooth closed
surface in the IIA solution defined by coordinate conditions such as those
above, then these conditions need not define a smooth closed surface
in the IIB solution.  This is exactly what happens in the gauge
used for $B$ in \cite{GM}, and performing the associated integrals over
our surfaces gives a nonvanishing result for $S_U$ and $S_0$.  The
sphere $S_L$ does not intersect these strings and so the corresponding
flux vanishes.  However, due to the Dirac string of the Kaluza-Klein
monopole the conditions (\ref{usphere},\ref{0sphere}) 
considered in the IIB solution in this gauge do not in fact define closed 
surfaces.

As a result, more care must be taken.  The simplest way to proceed
is to perform a change of gauge in the IIA solution
to move the Dirac strings of the B
field out of the way.  
This of course affects neither $C_1$ nor $C_3$.  These
Dirac strings can in fact be moved to $x_5 = x_6=0$ and into the region $v< b$
(i.e., inside the paraboloid).  To see this, we need only change to $v,w$ coordinates
in which the surface $v=b$ becomes a plane. 
The region $x_5=x_6=0, v < b$ then becomes the usual half-space shaped
Dirac string.  In this gauge, the strings do not intersect any of our
surfaces $S_U$, $S_0$, or $S_L$.  

Thus, using this new IIA gauge, all three surfaces do in fact
represent the closed surfaces we desire in the IIB solution.  In addition, the
total flux through any of these surfaces vanishes since they intersect no
Dirac strings.  We conclude that the
only D5-branes in the IIB solution are those lying at $v=w=0.$  

\section{Discussion}
\label{disc}

We have studied a situation in which, instead of transforming into
D5-brane charge, D4-brane charge simply disappears under T-duality.
Note that in the limit that the IIA NS5-brane degenerates to a line, the
IIA configuration appears to be just a D4-brane ending on a D6-brane.
On the other hand, the T-dual solution is a D5-brane attached to
a strange degenerate configuration of Kaluza-Klein monopole.  This latter
is some sort of singularity that carries no charge.

There is clearly some tension with the usual result that the T-dual of
a brane-ending configuration is another brane-ending configuration or
perhaps a three-brane junction.  The resolution appears to stem from
the fact that the IIA configuration is a rather 
subtle type of brane-ending solution.
This can be seen by comparing our case with the worldvolume solution of
\cite{Aki1} representing a fundamental string ending on a D2-brane.
One important difference is that in the case of \cite{Aki1} brane-source charge
is conserved and flows through the D2-brane from the junction out
to infinity.  In contrast, it was explicitly shown in \cite{GM} that the
D4-brane brane-source charge is not conserved in our IIA setting and is 
directly absorbed by the D6-brane.  Indeed, had D4-charge been conserved
on the IIA side then it would have been consistent to match this charge
to the D5-brane charge on the IIB side (which is certainly conserved).
In addition, had D4-brane charge flowed
through the D6-brane on the IIA side, one would certainly expect the 
IIB solution to replace this D6-brane with D5-branes of two sorts: one sort
coming from the D4-branes and one from the D6-branes.  It would then
be quite natural for the IIB solution to be a three-brane junction.

The fact that our solution is not of this type has to do with a somewhat
confusing 
property of D6-branes.  Consider any massless type IIA solution containing
D6-branes.  This of course provides a solution to 11-dimensional
supergravity in which the D6-branes are replaced by the cores of Kaluza-Klein
monopoles.  This solution has a Killing field $\lambda^{11}$ which vanishes
at the core of each monopole.  
The natural boundary condition to impose on the D6-branes is that
the corresponding 11-dimensional solutions be smooth\footnote{In the
case of unit-charged monopoles.  In the
case of multiply-charged Kaluza-Klein monopoles, the natural boundary
condition is that an appropriate multiple
cover be smooth.} at these cores.
But now consider the 11-dimensional four-form field  strength
$F_4^{\{11\}}$.  If it is smooth then $F_4^{\{11\}} \cdot 
\lambda^{11}$ must vanish when $\lambda^{11}$ does and in particular at
any core.  Since $H_3 = F_4^{\{11\}} \cdot \lambda^{11}$, it follows that
$H_3$ will vanish
at any D6-brane.  Note that since
the lowest Fourier mode around the circle will again give some smooth
field, this conclusion also holds in cases
where the 11-dimensional solution does not have an exact translation
symmetry along $\lambda_{11}$ but which can be treated perturbatively.  
The same argument also applies to the dual field, so that
$*\widetilde{F}_4=*_{11}F_{4}^{\{ 11 \} }$ should also vanish at a D6-brane.  
Here $*_{11}$ is the eleven-dimensional Hodge dual.  This is
the flux that causes D6-branes to produce fundamental strings, so 
no fundamental string
charge should be induced on a D6-brane when a D2-brane is dragged past it.

A similar sort of flux-excluding property was
studied in \cite{sup}.  For the `superconducting'
branes considered in that work, the normal component of some field
strength was forced to vanish on the horizon.  The situation here
is somewhat different, however, as now
the entire field strength $H_3$ or $*\widetilde{F}_4$ must vanish at the brane.  

The unexpected nature of our IIA solution can be directly traced to
this property\footnote{As can the fact that 
the D6-brane did not
deform in space as the NS5-brane was pulled past it, since 
the deformation of a brane is associated \cite{half} with
the tension from the induced charge flowing through the brane
to infinity.}.  Consider again the four stages of section \ref{shn}.
If $H_3$ had not vanished at the D6-brane, then the
D6-brane would have captured some non-zero flux from the NS5-brane.
This would in turn have resulted in the creation of D4-brane charge
at the center of the D6-brane which would then flow along the D6-brane
to infinity.  In stages (ii) and (iii) we would see D4-brane charge
absorbed on the NS5-brane and created on the D6-brane.
However, in the limiting case (iv) the charge would in fact have been conserved
as the absorption and creation would both occur at the same place 
(the intersection).  As a result, in this alternate setting one could have
said at stage (iv) that the D4 charge flows directly from 
the NS5-brane to the D6-brane.

Although we have identified the property that leads to our unexpected results
and we have shown that this property is a natural result of the eleven-dimensional
description, certain aspects of this story remain quite puzzling.
For example, flux is clearly
not expelled from the NS5-brane or from a corresponding D2-brane crossing
a D6-brane.  Yet, the D6-brane is connected to these other branes by dualities.
Thus, at least naively it appears that application of supergravity dualities
can transform the solutions of \cite{GM} into ones in which D6-branes
do in fact admit flux from other branes.  On the other hand, suppose that these
dualities do in fact produce a solution in which the D6-brane admits flux,
D4-brane charge flows into the D6-brane at stage (iv), and the IIB 
solution dual to stage (iv) is a three D5-brane junction.  What then
will be the IIB story at stages (ii) and (iii) where the branes are not
in contact?  Recall that it was in fact the flux-expulsion property of
our D6-brane that led to the resolution described in section \ref{IIB}.
This line of thinking suggests that some subtlety must arise in dualizing the
solutions of \cite{GM} in this way.  

Another avenue that should be explored is to
compare the IIB solution discussed here with recent solutions
\cite{FS1,FS2,FS3} representing 
M5-brane junctions and other holomorphic curves of M5-branes.
Under an appropriate set
of operations one should be able to transform these solutions into
three-brane junctions and brane-terminating solutions similar to the ones
discussed here.  Now, the solutions of \cite{FS1,FS2,FS3}
have certain odd properties of their own.  Consider for example \cite{FS1}, 
which analyzes an intersection two M5-branes in which  
the two M5-branes should in principle play symmetric roles.  Now, the solution
found in \cite{FS1}  is valid only in the near-horizon limit of one of the branes.
While this does break the explicit symmetry between the branes, the fact
that it is supposed to be part of a more complete symmetric solution implies
that certain properties of the two branes should be the same.  Yet, 
a close examination of the solution 
shows that one brane lies at a singularity while the other has a smooth horizon.
The second (smooth) brane is the one we are `near' in the solution and here the 
spacetime is much like AdS space.
Performing a sufficient number of dualities and
comparing the result with the solutions of \cite{GM} might therefore
yield insight into the puzzles associated with both solutions.

In summary, it would appear that two sorts of brane-terminating solutions
occur.  In the first, the termination occurs on a brane that admits flux
from other branes.  Here the charge of the terminated brane is conserved and flows
through the other brane back out to infinity.  T-dualizing such solutions leads
to further brane-ending solutions or to three-brane junctions.  In the
second sort, the termination occurs on a brane that for some reason
does not capture the proper sort of flux and the charge of the terminating
brane evaporates when it contacts the second brane.  In this latter
case T-duality leads to a solution with charge only at the location of this
second brane, the terminating brane having been replaced only with
a singularity that carries no charge though
it may carry multipole moments under various fields.  Whether these two
sorts of intersections are connected by dualities remains unclear.


\acknowledgments

The author would like to thank  Neil R. Constable, Ben Craps, Gary Gibbons,
Andr\'es Gomberoff, Jeff Harvey, Clifford Johnson, Emil
Martinec, Rob Myers, Amanda Peet, 
Savdeep Sethi, and Oyvind Tafjord
for useful discussions.  This work was supported in part by
NSF grant PHY97-22362 to Syracuse University,
the Alfred P. Sloan foundation, and by funds from Syracuse
University.  


\end{document}